\documentclass[11pt]{article}
\usepackage{graphicx}
\usepackage{amssymb,amsfonts,amsmath}
\usepackage{amsbsy}
\usepackage{setspace}

\begin{document}
\title{A boundary integral formalism for stochastic ray tracing in billiards}
\author{David J. Chappell$^{1}$ and Gregor Tanner$^{2}$ \\
1. School of Science and Technology,\\ Nottingham Trent University, Clifton Campus,\\ Nottingham NG11 8NS,  UK.\\
2. School of Mathematical Sciences,\\ University of Nottingham,
University Park,\\ Nottingham NG7 2RD, UK.}
\date{}
\maketitle

\abstract{Determining the flow of rays or particles driven by a
force or velocity field is fundamental to modelling many physical
processes, including weather forecasting and the simulation of
molecular dynamics. High frequency wave energy distributions can
also be approximated using flow or transport equations. Applications
arise in underwater and room acoustics, vibro-acoustics, seismology,
electromagnetics, quantum mechanics and in producing computer
generated imagery. In many practical applications, the driving field
is not known exactly and the dynamics are determined only up to a
degree of uncertainty. This paper presents a boundary integral
framework for propagating flows including uncertainties, which is
shown to systematically interpolate between a deterministic and a
completely random description of the trajectory propagation. A
simple but efficient discretisation approach is applied to model
uncertain billiard
dynamics in an integrable rectangular domain.}\\

\noindent\textbf{Many physical transport problems can be formulated
in terms of ray tracing or trajectory methods. Applications range
from particle tracking in fluids \cite{CCMM04, SR10} and the
simulation of molecular dynamics \cite{Noe09} to illumination and
rendering problems in computer graphics \cite{JK86} or, more
generally, the geometric optics limit of linear wave equations.  A
range of techniques have been developed for solving ray tracing
problems. One distinguishes between direct ray-tracing \cite{Vor89,
Cer01} based on following ray paths from a source to receiver point
and variants thereof; and indirect methods using transport equations
based on conservation laws such as the Liouville equation
\cite{LeV92} to propagate phase space densities. In the latter case
one arrives at a model for propagating phase-space densities using
deterministic transfer operators of the Frobenius-Perron (FP) type
\cite{Cvi12}. In this paper we will introduce a new boundary
integral method for determining phase-space densities propagated via
a stochastic trajectory flow using a transfer operator approach.}\\

\section{Introduction}

A variety of techniques have emerged recently with the aim of
turning transfer operators into an efficient numerical tool for
practical applications. Domain based transfer operator approaches,
for example, start by subdividing the phase-space into distinct
cells and considering transition rates between these phase-space
regions. One of the simplest and most common approaches of this type
is Ulam's method (see e.g.\ \cite{JD96}). Other methods include
wavelet and spectral methods for the infinitesimal FP-operator
\cite{JK09, FJK11}, eigenfunction expansion methods \cite{BMM12} and
periodic orbit expansion techniques \cite{Cvi12, PC10}. Also the
modelling of many-particle dynamics, such as protein folding, has
been approached using short trajectories of the full,
high-dimensional molecular dynamics simulation to construct reduced
Markov models \cite{Noe09}. For a discussion of convergence
properties of the Ulam method in one and several dimensions, see
\cite{BM01} and \cite{BKL02}, respectively. However, such methods
have only found a fairly limited range of applicability, with
difficulties arising due to the high-dimensionality of the
phase-space.

In the following we will focus on integral equation formulations for
propagating phase space densities along ray trajectories using
transfer operators. One such formulation is given by the rendering
equation \cite{JK86} which has its origins in computer graphics, but
has been applied more widely since \cite{AL06, GT09}. The rendering
equation can again be formulated in terms of transfer operators
\cite{GT09, SAE14}. A boundary integral FP-operator approach called
{\em dynamical energy analysis} (DEA) has been introduced in
\cite{GT09} and further developed in \cite{CGT11}. In a sequence of
papers \cite{CTG12, CT13} the method has evolved into an mesh-based
tool called {\em discrete flow mapping} (DFM) described in
\cite{PRSL13, Wamot14}. This has proven to be an efficient numerical
tool making it possible to handle trajectory flow problems on
complex surfaces (consisting of circa $10^5$ to $10^6$ mesh cells)
on the time-scale of a few hours on standard desktop computers
\cite{Wamot14}.

Here we will extend the DEA approach towards dynamical systems with
uncertainties and stochastic dynamics. The reasons for doing so are
twofold: firstly, in many physically relevant situations, the system
dynamics are inherently stochastic or system parameters are not
known exactly and a probabilistic approach will be necessary.
Secondly, including stochasticity in a transfer operator changes the
properties of the operator fundamentally in a way that opens the
door for a wider range of numerical solution techniques. Techniques
for constructing stochastic ray-tracing operators have been
presented in \cite{PC98, PC99a, PC99b, GP01} in the context of the
FP operator, and in acoustics in terms of the radiosity equation
\cite{AL98}.

In this paper we construct a stochastic ray-tracing
operator that leads to a boundary integral formulation for
stochastic dynamics in billiards.
That is, the underlying dynamical system is that of a particle or point mass
moving on a billiard table with constant velocity (without friction) inside a compact domain
$\Omega$ with piecewise smooth boundary $\Gamma$ as described by
Sinai \cite{YS04}, see also \cite{Cvi12}, Ch. I, Sec. 8. The particle
is assumed to undergo specular
reflections upon collision with the smooth sections of $\Gamma$. As
the overall energy of the system is constant, the billiard dynamics
(integrable, mixed or chaotic) is completely controlled by the
geometry of $\Gamma$. However, for the stochastic evolution
considered here, both the position of the transported particle and
the nature of its reflection at the boundary will be considered as
uncertain. Typically, the mean transported position and reflected
direction will be those of the standard (deterministic) billiard
map. The effect is that total energy remains constant, but the
stochasticity will clearly influence the billiard dynamics as will
be explored in Section \ref{Sec:dynsys}.
The resulting stochastic evolution operator will be of Fokker-Planck type
as discussed in \cite{PC10, PC98}.

We note that statistical methods related to the stochastic approach
proposed here have been used in a variety of engineering
applications. In particular, the so-called statistical energy
analysis (SEA) (see for example \cite{RL69} and \cite{RL95}) for
modelling vibro-acoustic energy distributions and the random
coupling model (RCM) \cite{RCM} for modelling electromagnetic
fields, see also \cite{Gradoni14}. In SEA and RCM the structure is
subdivided into a set of subsystems and ergodicity of the underlying
ray dynamics as well as quasi-equilibrium conditions are postulated.
The result is that the density in each subsystem is taken to be
approximately constant leading to greatly simplified equations based
only on coupling constants between subsystems. The disadvantage of
these methods is that the underlying assumptions are often hard to
verify {\em a priori} or are only justified when an additional
averaging over `equivalent' subsystems is considered. The
shortcomings of SEA have been addressed by Langley \cite{RL92, RL94}
and more recently in a series of papers by Le Bot \cite{AL06, AL98,
AL02}.

In this work we focus on stochastic ray-tracing approximations for
linear wave problems in two-dimensions, or equivalently on
stochastic billiard dynamics; the models developed can easily be
generalized to higher dimensions. We propose a new boundary integral
approach based on the use of stochastic evolution operators to
incorporate uncertain ray dynamics into our model in a quantifiable
manner. Propagating densities with uniformly distributed probability
of location and direction leads to the quasi-equilibrium approaches
mentioned above (SEA and RCM). We will show that choosing a scaled
and truncated Gaussian probability distribution instead leads to a
model that interpolates between SEA and deterministic ray tracing.
This interpolation takes place at the level of the governing model,
in contrast to DEA which provides a similar interpolation due to the
precision of the chosen numerical approximation method \cite{GT09}.
Once an estimate of the level of uncertainty in the model has been
prescribed, an appropriate numerical solution approach can be
applied.

The paper is structured as follows: in Sec.\ \ref{sec:int-eq} a
boundary integral description of deterministic ray tracing in
billiards will be presented. The addition of noise into the model
will then be outlined and an approach that interpolates between a
deterministic and a random trajectory flow will be described. In
Sec.\ \ref{sec:impl}, the numerical implementation of the model will
be outlined and illustrated via the example of stochastic ray
tracing in a rectangular billiard. The decay of correlations and the
asymptotic escape rate will be studied to diagnose the behaviour of
the rectangular billiard model as it makes the transition from
regular and deterministic to probabilistic dynamics.

\section{Boundary Integral Equation Formulation}
\label{sec:int-eq}

\subsection{A boundary integral description of deterministic ray tracing via transfer operators}\label{sec:detop}
\begin{figure}
\begin{center}
\includegraphics[width=0.8\textwidth]{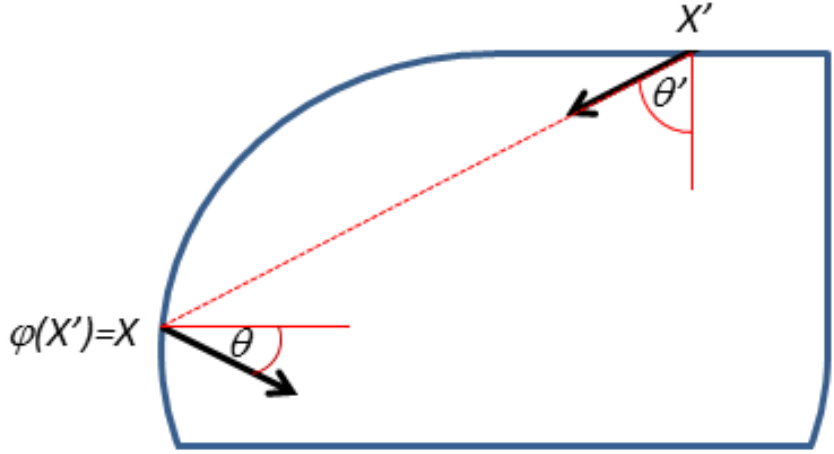}
\end{center}
\caption{Propagation of ray trajectories using a deterministic
boundary map.} \label{boundary-map}
\end{figure}

Consider the trajectory flow described by a Hamiltonian
$\hat{H}=c|\mathbf{p}|$ in a finite two-dimensional domain $\Omega$
as depicted in Fig.\ \ref{boundary-map}, where $c$ is the speed of
propagation and $\mathbf{p}$ is the inward momentum (or slowness)
vector. Denote the phase-space on the boundary of $\Omega$ with
fixed total energy $\hat{H}=1$ as $Q =
\Gamma\times(-c^{-1},c^{-1})$, where $\Gamma$ is the boundary of
$\Omega$. The associated coordinates are $X=[s,\:p] \in Q$ with
$s\in[0,L)$ (arc-length) parameterising $\Gamma$ and  $p \in
(-c^{-1},c^{-1})$ parameterising the component $\mathbf{p}$
tangential to $\Gamma$. Explicitly, the momentum coordinate $p$ is
defined in terms of the angle $\theta$ between $\mathbf{p}$ and the
normal to $\Gamma$ at $s$ (see Fig.\ \ref{boundary-map}) as
$p=c^{-1}\sin(\theta)$. We adopt the convention that $\theta
\in(-\pi/2,\pi/2)$ and is positive for counter-clockwise
propagation. The deterministic boundary flow map is denoted
$\varphi:Q\rightarrow Q$, and maps a vector in $Q$ via the
Hamiltonian flow to another vector in a subset of $Q$. This map
defines a deterministic evolution of the form $\varphi(X')=X_{}$,
where $X'_{}=[s',\:p']$, $X=[s,p]$ . Fig.\ \ref{boundary-map} shows
that geometrically $\varphi$ corresponds to the composition of a
translation (from $s'$ to $s_{}$) and a rotation to the direction
corresponding to a specular reflection.

The propagation of a phase-space density $\rho$ by the boundary map
$\varphi$ through a single reflection is given by the
Frobenius-Perron operator acting on this map
\begin{equation}\label{FPO}
\mathcal{L}\rho(X)=\int_{Q}\delta(X-\varphi(X'))\rho(X')dX'.
\end{equation}
For an initial boundary distribution $\rho_{0}$ on $Q$, the final
density after adding contributions from all refections may be
computed using the following boundary integral equation (see
\cite{GT09}, \cite{CTG12} and \cite {CT13}),
\begin{equation}\label{DetIE}
(I-\mathcal{L})\rho=\rho_{0}.
\end{equation}
Note that for the sum over all reflections to converge, energy
losses must be introduced into the system, which could take place at
the boundaries themselves, or along the trajectories. In general, a
weight factor $w$ will be added inside the integral in the
definition of $\mathcal{L}$ which contains a dissipative term, and
for the extension to multiple domains connected at interfaces $w$
will also contain reflection/transmission probabilities at these
interfaces. For non-convex polygons, $w$ will additionally include a
visibility function.

\subsection{Stochastic trajectory tracking in billiards}\label{sec:theory}
\subsubsection{The stochastic propagation operator}
Building upon the deterministic propagation models described in the
previous section, we propose a family of phase space density
propagation models with transfer operators of the form
\begin{equation}\label{StochasticTO}
\mathcal{L}_{\boldsymbol{\sigma}}\rho(X)=\int_Q
f_{\boldsymbol{\sigma}}(X-\varphi(X'))\rho(X')dX'.
\end{equation}
This operator bears a strong similarity to (\ref{FPO}), but the
$\delta$ distribution term has been replaced with a probability
density function (PDF) $f_{\boldsymbol{\sigma}}$ such that
\begin{equation}\label{StochasticPDF}
\int_Q f_{\boldsymbol{\sigma}}(X)dX=1.
\end{equation}
Here, $f_{\boldsymbol{\sigma}}$ is the probability distribution and
$\boldsymbol{\sigma}$ is the parameter set controlling its shape.
With reference to applications, such a probabilistic behaviour could
be attributed to, for example, fluctuations in the wave speed $c$,
roughness of the reflecting surface or uncertainty in the exact
position of the boundary. In the following, we will always assume
that the total energy $\hat{H}=c|\mathbf{p}| = 1$ remains fixed and
that the total probability is conserved, that is, condition
(\ref{StochasticPDF}) holds throughout. Note that in contrast to the
models considered in \cite{PC10, PC98}, the range of integration in
the billiard models considered here is in general bounded, which has
implications for the choice of suitable PDFs
$f_{\boldsymbol{\sigma}}$.

The simplest case is to take $f_{\boldsymbol{\sigma}} = const$, upon
which one arrives at a model describing propagation to all
admissible positions and directions with equal probability. The
systems is thus by definition ergodic and independent of the
underlying classical dynamics. Note that ergodicity is a key
assumption for an SEA or RCM treatment to be valid.

\begin{figure}
\begin{center}
\includegraphics[width=\textwidth]{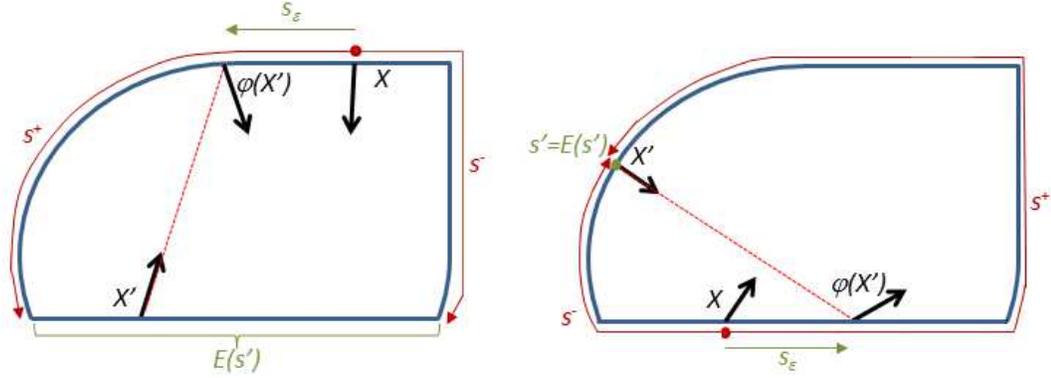}
\end{center}
\caption{Tracking ray trajectories via a noisy boundary map and
truncation limits $s^\pm$ for the random variable $s_\varepsilon$.}
\label{TruncRange}
\end{figure}

In general, we would like to arrive at a stochastic operator which
includes both the deterministic operator in Eq.\  (\ref{FPO}) and
the random propagation model described above as limiting cases. In
addition, the PDF $f_{\boldsymbol{\sigma}}$ needs to obey conditions
on the sampling ranges due to the limited range of the boundary map
$\varphi$. For simplicity we will restrict to convex domains
$\Omega$ to avoid additional complications due to incorporating
visibility functions.

\subsubsection{The probability density function - normalisation}
We may interpret the evolution
given by the operator in Eq.\ (\ref{StochasticTO}) as originating from a
stochastic boundary map $\varphi_{\boldsymbol{\sigma}}$ with added noise,
that is,
\begin{align}\label{Gaussian}
\begin{split}
\varphi_{\boldsymbol{\sigma}}(X') &= {X},\\
&=\varphi(X')+X_\varepsilon,
\end{split}
\end{align}
where $X_\varepsilon = [s_\varepsilon, p_\varepsilon]$ are random
variables drawn from the  PDF $f_{\boldsymbol{\sigma}}$. Note that
$s_\varepsilon$ is understood as a shift in counter-clockwise
direction. For $ X \in Q$ given, we have to ensure that
$\varphi(X')= X - X_\varepsilon$ is still in the range of the
deterministic map $\varphi$; this yields restrictions on the
possible values of $X_\varepsilon$ and thus on the domain of
$f_{\boldsymbol{\sigma}}$.

We define $\varphi=[\varphi_s,\:\varphi_p]$ in terms of its position
and momentum components and write the initial coordinate as $X' =
[s',p']$. The range of admissible values for $\varphi_s(X')$ is
$[0,L)\setminus E(s')$, where $E(s')$ is the (closed) set of all
points on the same straight edge as $s'$, see Fig.\
\ref{TruncRange}. Note that for curved edges we set $E(s')=s'$ as
shown on the RHS of Fig.\ \ref{TruncRange}. Furthermore, we have
that $\varphi_p(X')\in(-c^{-1},c^{-1})$. It is therefore necessary
to truncate the ranges from which $X_\varepsilon$ are sampled to the
ranges where for fixed $X$, $\varphi(X')\in([0,L)\setminus
E(s'))\times(-c^{-1},c^{-1})$ in Eq.\ (\ref{Gaussian}). Denoting
these truncated ranges by $(X^-,X^+)$ where
$X^{\pm}=[s^\pm,\:p^\pm]$, the PDF $f_{\boldsymbol{ \sigma}}$ will
have support on $X_\varepsilon\in(X^-,X^+)$ only. The truncated
sampling ranges are given as $s^+(s',s)=\min\{x>0: s+x\in
{E}(s')\:\:(\mathrm{mod}\:L)\}$ and correspondingly
$s^-(s',s)=\max\{x<0:s+x\in {E}(s')\:\:(\mathrm{mod}\:L)\}$ (see
Fig.\ \ref{TruncRange}). Likewise in the momentum coordinate,
$p^+(p)=c^{-1}-p$ and $p^-(p)=-c^{-1}-p$. Using Heaviside functions
we define a cut-off function for restricting the support of
$f_{\boldsymbol{\sigma}}$ to $(X^-,X^+)$ as follows
\begin{align}
\begin{split}
&\chi(X_\varepsilon;X^-,X^+)\\
&=(H(s^+-s_\varepsilon)-H(s^- -s_\varepsilon))(H(p^+-p_\varepsilon)-H(p^--p_\varepsilon)).\\
\end{split}
\end{align}
Note that we have omitted the dependence of $s^\pm$ and $p^\pm$ on
$X'$ and $X$ for brevity.

Having obtained the domain of the PDF, we can now construct
$f_{\boldsymbol{\sigma}}$ explicitly; we will derive the PDF from an
uncorrelated bivariate Gaussian distribution with mean
$\mathbf{0}=[0,0]$ and standard deviation $\boldsymbol{\sigma}=
[\sigma_1,\sigma_2]$. A normalized PDF is then obtained by setting
\begin{align}\label{ScaledPDF}
f_{\boldsymbol{\sigma}}(X_\varepsilon;X^-,X^+)=\frac{\chi(X_\varepsilon;X^-,X^+)\exp{\left(\displaystyle-\frac{s_\varepsilon^2}{2\sigma_1^2}\right)}\exp{\left(\displaystyle-\frac{p_\varepsilon^2}{2\sigma_2^2}\right)}}{2\pi\sigma_1\sigma_2\psi_{\sigma_1}(s^-,s^+)\psi_{\sigma_2}(p^-,p^+)},
\end{align}
where the normalization defined through $\psi_{\sigma_1}$ and $
\psi_{\sigma_2}$ is given as
\begin{align}
\psi_{\sigma_1}(s^-,s^+)=\frac{1}{2}\left(\mathrm{erf}\left(\frac{s^+}{\sqrt{2}\sigma_1}\right)-\mathrm{erf}\left(\frac{s^-}{\sqrt{2}\sigma_1}\right)\right),
\end{align}
and $\psi_{\sigma_2}$ is defined analogously. The normalisation
ensures that the PDF satisfies condition (\ref{StochasticPDF}) for
the truncated sampling ranges specified through $\chi$. Note that
the mean and variance of $f_{\boldsymbol{\sigma}}$ differs in
general from that of the underlying Gaussian distribution.

The two limiting PDFs are obtained by considering the limiting
values of ${\boldsymbol{\sigma}}$. Taking the limit of
(\ref{ScaledPDF}) as $\boldsymbol{\sigma}\rightarrow\mathbf{0}$ then
\begin{equation}\label{PDF0}
f_{\boldsymbol{\sigma}}(X_\varepsilon;X^-,X^+)\rightarrow\lim_{\boldsymbol{\sigma}\rightarrow\mathbf{0}}\frac{\chi(X_\varepsilon;X^-,X^+)}{2\pi\sigma_1\sigma_2}\exp{\left(\displaystyle-\frac{s_\varepsilon^2}{2\sigma_1^2}\right)}\exp{\left(\displaystyle-\frac{p_\varepsilon^2}{2\sigma_2^2}\right)}.
\end{equation}
The distribution becomes increasingly sharp and the bivariate
Gaussian tends to a two-dimensional delta distribution localised
around $X_\varepsilon = X- \varphi(X') = 0$, which describes the
deterministic flow discussed in the Section \ref{sec:detop}. Taking
the limit as $\sigma_1$, and $\sigma_2$ go to $\infty$ and using the
leading order asymptotic expansion of the error function about 0
returns
\begin{equation}\label{PDFinf}
f_{\boldsymbol{\sigma}}(X_\varepsilon;X^-,X^+)\rightarrow\frac{c}{2(s^+-s^-)}\chi(X_\varepsilon;X^-,X^+).
\end{equation}
Note that this is just the uniform distribution for
$s_\varepsilon\in(s^-,s^+)$ and $p_\varepsilon\in(p^-,p^+)$ (since
$p^+-p^-=2c^{-1}$) leading to the fully probabilistic regime
described above. The mean and variance of the normalized
distribution may be calculated from the PDF (\ref{ScaledPDF}) using
the standard formulae. The variance of the bivariate distribution
will tend to $\boldsymbol{\sigma}$ as
$\boldsymbol{\sigma}\rightarrow\mathbf{0}$. For large $\sigma_j$,
$j=1,2$ we have the variance of the uniform distribution. That is,
as $\sigma_1\rightarrow\infty$,
$\mathrm{Var}(s_{\varepsilon})=(s^+-s^-)^2/12$, and as
$\sigma_2\rightarrow\infty$, then
$\mathrm{Var}(p_{\varepsilon})=1/(3c^2)$. Clearly such data are
vital for applications in uncertainty quantification, for example,
for modelling uncertain high frequency vibro-acoustic or
electromagnetic wave propagation through a manufactured structure or
device.

We turn our attention to propagating a density along stochastic ray
paths according to the PDF (\ref{ScaledPDF}) via the transfer operator (\ref{StochasticTO}). We
proceed by considering the numerical evaluation of
$\mathcal{L}_{\boldsymbol{\sigma}}$; we will in particular consider some important
dynamical quantities, namely the rate of escape and the
decay of correlations. These will be studied to help diagnose the
behaviour of the model for different ranges of $\boldsymbol{\sigma}$.

\section{Implementation and Results}\label{sec:impl}
\subsection{Discretisation}
A number of efficient methods for evaluating $\mathcal{L}$
numerically in domains including complex multi-component systems
have recently been developed \cite{PRSL13, Wamot14}. One advantage
of instead working with $\mathcal{L_{\boldsymbol\sigma}}$ is that it
is a compact integral operator and hence may be evaluated more
simply via direct discretisation methods rather than the variational
approaches described in \cite{PRSL13, Wamot14} and references
therein.
\begin{figure}
\begin{center}
\includegraphics[width=0.75\textwidth]{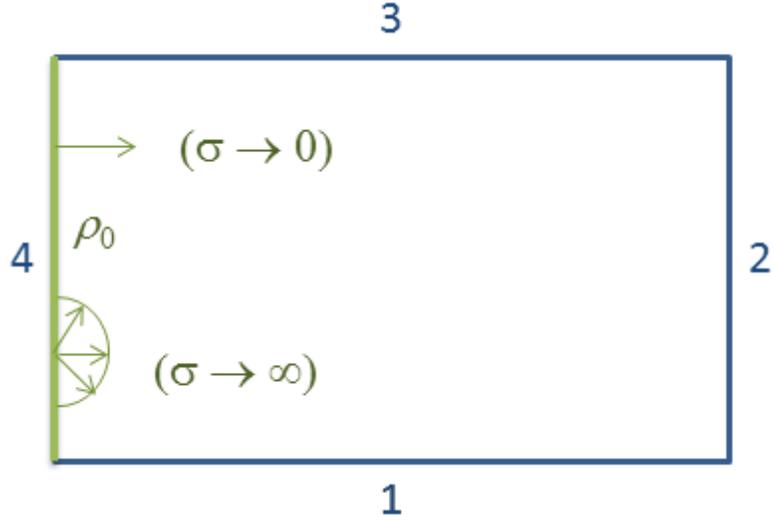}
\end{center}
\caption{A rectangular billiard with prescribed boundary condition
$\rho_0$.} \label{Rect}
\end{figure}

Here we approximate $\mathcal{L}_{\boldsymbol\sigma}$ on a
rectangular billiard as shown in Fig. \ref{Rect}. The reason for
choosing this simple domain is that its integrable dynamics make it
ideal for identifying the effect of varying $\boldsymbol\sigma$ in
isolation of other sources of ray chaotic behaviour. In particular,
we make use of our experience in dealing with domains with corners
in \cite{PRSL13, Wamot14} and employ a piecewise constant
collocation method with $n$ elements in the position variable,
collocating at element centers. That is, we separate out and
approximate the spatial dependence of $\rho$ in the form
\begin{equation}
\rho(s,p)\approx\tilde{\rho}(p)\sum_{j=1}^{n}a_jb_j(s),
\end{equation}
where $b_j(s)=1$ if $s$ lies on the $j$th element and zero
elsewhere. The coefficients $a_j$ are the unknowns to be determined.
The semi-discrete operator $\mathcal{L}_\sigma$ is then evaluated at
the collocation points $s=s_i$ for $i=1,\ldots,n$ using equation
(\ref{StochasticTO}) as
\begin{equation}\label{StochasticTOdisc1}
\mathcal{L}_{\boldsymbol{\sigma}}\rho(s_i,p)=\sum_{j=1}^na_j\int_{-c^{-1}}^{c^{-1}}\tilde{\rho}(p')\int_{e_j}
f_{\boldsymbol{\sigma}}(X_i-\varphi(X'))ds'dp',
\end{equation}
where $X_i=[s_i,\: p]$ and the range of integration with respect to
$s'$ is on the $j$th element $e_j$. The phase space coordinate
$X'=[s',\:p']$ provides the variables integration $s'\in e_j$ and
$p'\in(-c^{-1},c^{-1})$. Note that the integral with respect to $s'$
may be calculated analytically in terms of the error function for
discretisation by flat (straight line) elements and using the
normalised PDFs described in the last section. This step is
important for \emph{efficient} computations of the discretised
transfer operator.

A full discretisation is then achieved by applying the Nystr\"{o}m
method in momentum space with $N$-point trapezoidal integration and
a step size $h$. Note that in order to evenly discretize with
respect to the direction of ray propagation, the integration
variable is changed from $p'$ to $\theta'$ using the relation
$cp'=\sin(\theta')$. This reduces the calculation in
(\ref{StochasticTOdisc1}) to a matrix-vector multiplication, with
matrix entries of the form
\begin{equation}\label{StochasticTOdisc2}
L_{I,J}=\frac{h\cos(\theta')}{c}\int_{e_j}
f_{\boldsymbol{\sigma}}(X_I-\varphi(X_J'))ds'.
\end{equation}
Here $X_I=[s_i,\: p_{\iota}]$ and so $I$ is the multi-index
$(i,\:\iota)$, with $p_\iota$, $\iota=1,\ldots,N$ giving the values
of the momenta corresponding to the trapezoidal rule grid points.
Likewise, $X_J'=[s',\: p_{k}]$ and $J$ is the multi-index $(j,\:k)$,
where $s'\in e_j$ is the integration point and $p_k$, $k=1,\ldots,N$
runs over the trapezoidal rule grid points as before. The density
$\rho$ can (by extension) be considered as periodic in the momentum
variable since $\rho(s,c^{-1})=\rho(s,-c^{-1})=0$, and so the
semi-discretisation in momentum space should converge
super-algebraically for smooth initial data. The convergence
properties of the method overall are demonstrated in the next
section. A further major advantage of this combination of methods is
that the need for numerical integration methods is completely
avoided.

\subsection{Convergence}

\begin{figure}
\begin{center}
\includegraphics[width=\textwidth]{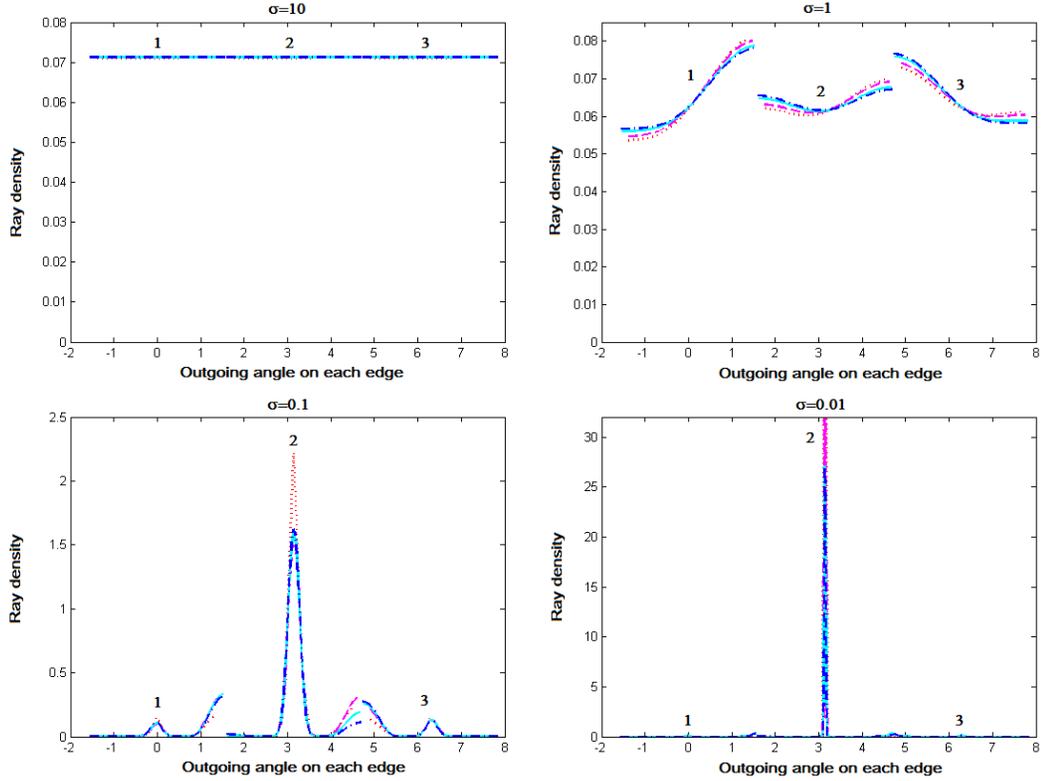}
\end{center}
\caption{Convergence of the ray density after one reflection and the
effect of changing $\sigma$ on the dynamics. For all plots except
$\sigma=0.01$: $\cdots$: $N=16$, $--$: $N=32$, ---: $N=64$,
$\cdot$-$\cdot$: $N=128$. For the $\sigma=0.01$, the
previous $N$ values should all be multiplied by 4. The horizontal
axis shows the outgoing angle in the range $-\pi/2$ to $\pi/2$ on
edge 1, on edge 2 it is shifted by $\pi$ and on edge 3 it is shifted
by $2\pi$. Edge numbers are indicated on the plot.}
\label{ConvSigma}
\end{figure}

To test the convergence of the approximation of
$\mathcal{L}_{\boldsymbol\sigma}$ we propagate a stochastic boundary
(line) source through a single reflection. The dimensions of the
rectangle are taken to be $0.75$ by $0.25$ and we let $c=1$ meaning
that both the position and momentum variables have the same total
range. We also take $\sigma_1=\sigma_2=\sigma$ for simplicity,
although the extension to distinct $\sigma_1$ and $\sigma_2$ is
clearly straightforward. We number the edges as shown in Fig.
\ref{Rect} so that edges $1$ and $3$ have length $0.75$ and take
\begin{equation}
\rho_0(s,p)=\frac{I_{4}(s)\exp\left(\displaystyle\frac{-p^2}{2\sigma^2}\right)}
{\sqrt{2\pi\sigma^2}\mathrm{erf}\left(\frac{1}{\sqrt{2}\sigma}\right)},
\end{equation}
where $I_4$ is an indicator function for edge number 4. That is, the
source is applied along edge 4 as shown in Fig. \ref{Rect} and its
directivity depends on the parameter $\sigma$. Figure
\ref{ConvSigma} shows the result of approximating
$\mathcal{L}_\sigma\rho_0$ on sides 1 to 3 of the rectangle. The
plot shows the mean ray density along each of the 3 edges plotted
against the outgoing angle. The horizontal axis is a shifted value
of this outgoing angle which is unshifted on side 1, shifted by
$\pi$ on side 2 and by $2\pi$ on side 3. This is simply to show the
results for each edge side-by-side.

Figure \ref{ConvSigma} shows the transition from probabilistic to
deterministic dynamics as $\sigma$ is decreased, and therefore
illustrates the theory outlined in the Section \ref{sec:theory}. In
particular, for $\sigma=10$ we see a uniformly distributed ray
density across all edges and all outgoing directions. For
$\sigma=0.01$ one sees that the ray density localises on edge 2 with
outgoing angle 0, i.e. perpendicular to the boundary. This is a
close approximation to the expected deterministic evolution. The
intermediate cases ($\sigma=1$ and $\sigma=0.1$) show the transition
between these two limiting cases. This transition will be considered
in more depth in the next section.
\begin{figure}
\begin{center}
\includegraphics[width=0.8\textwidth]{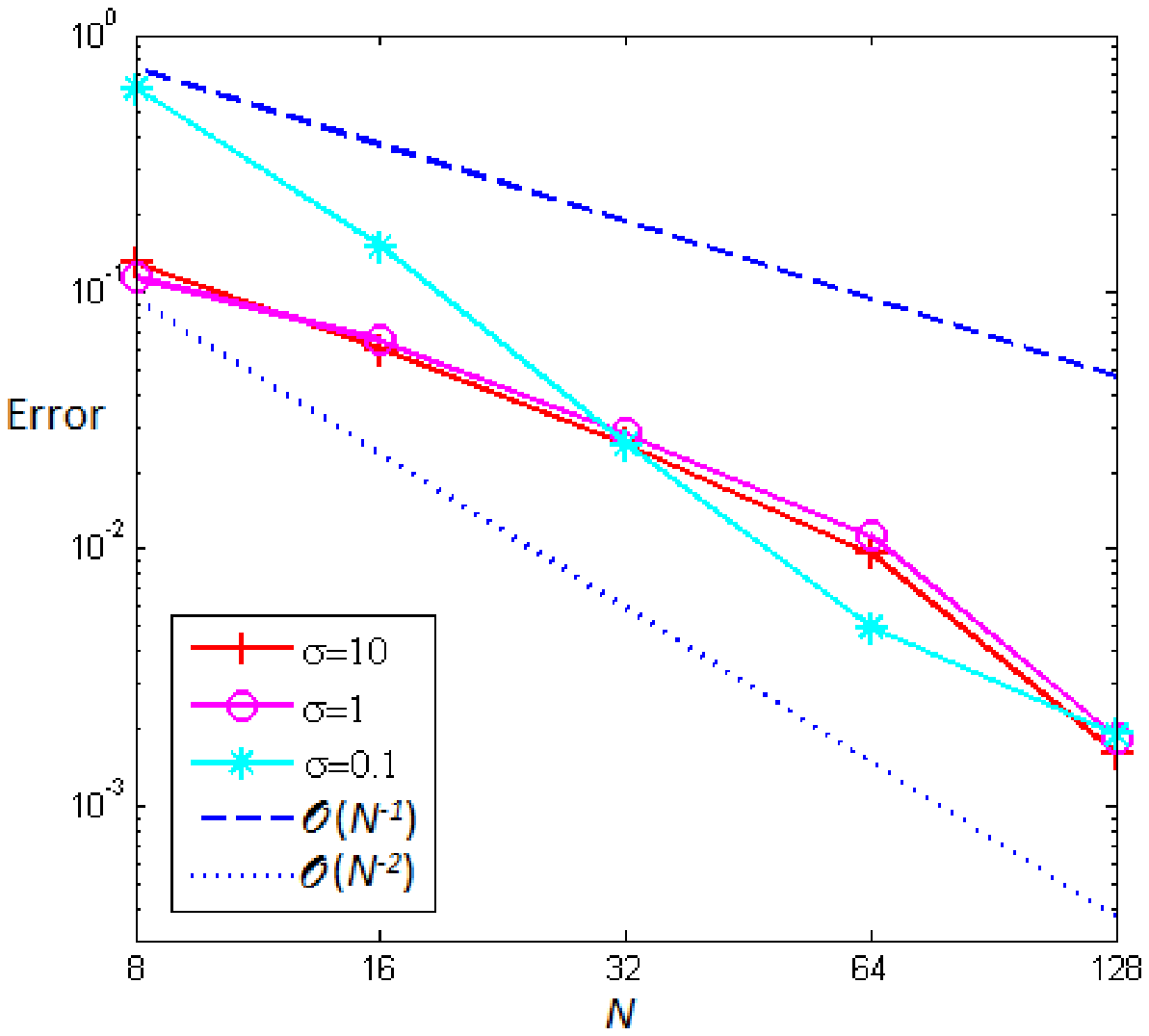}\end{center}
\caption{Convergence rate in computing the total ray density
(\ref{TRD}) for different values of $\sigma$.}
\label{ConvR}\end{figure}

In order to test the convergence of the results shown in Figure
\ref{ConvSigma}, we integrate the boundary phase-space density over
$Q$ to estimate the total density
\begin{equation}\label{TRD}
\rho_{\mathrm{tot}}=\int_Q\mathcal{L}_{\boldsymbol \sigma}\rho_0 dX.
\end{equation}
For the basic discretisation approaches employed here and taking
$n=N$ one typically sees convergence in computing
$\rho_{\mathrm{tot}}$ to the first few significant figures with
absolute errors of estimated order between $\mathcal{O}(N^{-1})$ and
$\mathcal{O}(N^{-2})$ as shown in Figure \ref{ConvR}. Note that
these rates appear to be superior to the sub-linear rates expected
from a standard Ulam approach \cite{BM01}. Convergence rates are
generally higher for smaller values of $\sigma$, and usually
increase slightly when the number of discretisation points $n=N$ in
both the position and momentum variables is increased. Note that for
$\sigma=0.01$, the method has only converged sufficiently to produce
meaningful results when $N\geqslant 128$ and as such this case has
been omitted from the figure. This suggests that the singularly
perturbed problem for small $\sigma$ should be tackled using an
adaptive meshing procedure to resolve the peak(s) more efficiently,
rather than the uniform grid employed here. The development and
analysis of such approaches will be considered as part of future
work.

\subsection{Rate of escape and decay of
correlations}\label{Sec:dynsys}
The rates of escape and decay of correlation provide useful
information about the dynamics of the billiard system being studied
in terms of their description and classification (chaotic, mixed or
integrable).
The escape rate $\gamma$ measures  the decay
of the total phase space density, that is, the survival probability,
in case of an open or absorbing billiard.  This decay is exponential
for chaotic dynamics, that is,
\[ \int_Q \left[{\cal L}^n_{\sigma} \rho_0\right](X) \;dX \sim e^{-\gamma n};\]
similarly, for closed, chaotic systems, the decay of correlation scales
exponentially with a decay rate $\nu$ according to
\[\int_Q \rho_0 (X)  \left[{\cal L}^n_{\sigma}\rho_0\right](X)\; dX -
\left[\int_Q \rho_0 (X)  \overline{\rho}(X)\; dX\right]^2
\sim e^{-\nu n},\]
where  $\overline{\rho} = \lim_{n\to\infty} {\cal L}_{\sigma}^n \rho_0$ is the
natural density (if the limit exists).
Both, $\gamma$ and $\nu$ are closely linked to the spectrum of
${\cal L_{\sigma}}$ with $\exp(-\gamma)$ and $\exp(-\nu)$ being the
magnitude of the leading and next-leading eigenvalue of ${\cal L_{\sigma}}$ for
ergodic dynamics \cite{Cvi12}.

The rates $\gamma, \nu$ are also important when considering wave energy
propagation through a built up structure \cite{CT11}. In particular,
the suitability of the random wave superposition hypothesise of an
SEA-type approach can be analysed in this framework, since a fast
decay of correlations compared to the escape rate provide the ideal
setting for a diffuse random wave field to be created \cite{GT09}.
On the other hand, slow or non-decaying correlations in the dynamics
indicate regularity in the wave field and will
introduce non-random fluctuations and potentially long range
correlations between multiple sub-domains.

In this section we study the decay of correlations in the
rectangular billiard described earlier for different choices of the
parameter $\sigma$. In addition, we consider the rate of escape when
a small opening is introduced on the boundary and consider the
effect of changing both $\sigma$ and the size of the opening.

\begin{figure}
\begin{center}
\includegraphics[width=\textwidth]{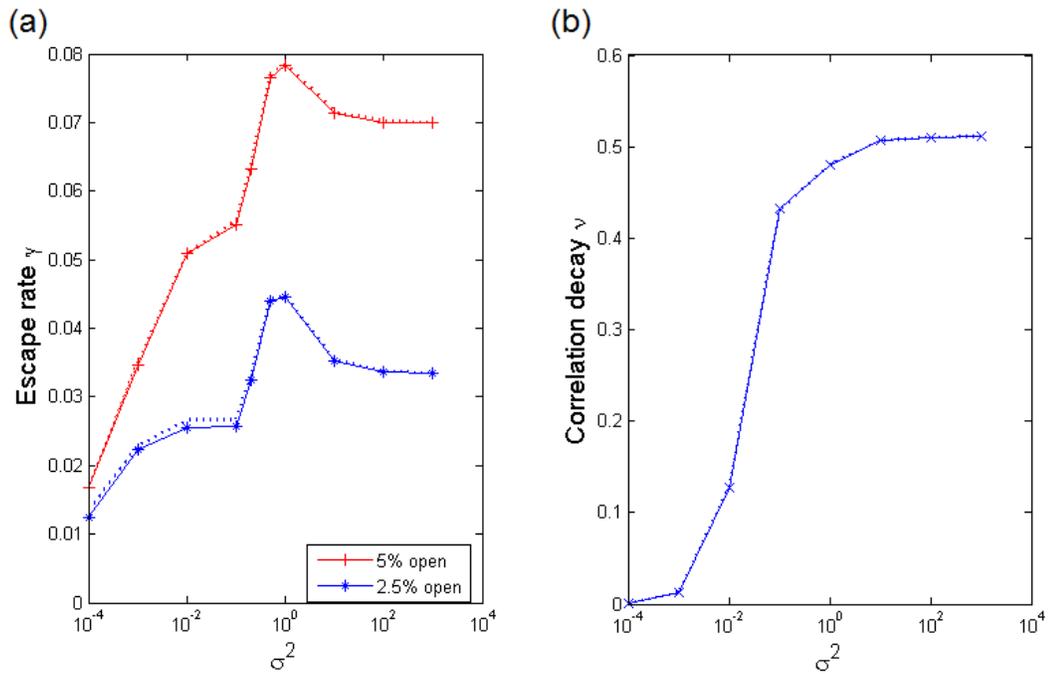}
\end{center}
\caption{(a) The dependence of the asymptotic escape rate on
$\sigma^2$ in a rectangular billiard for two different hole sizes.
(b) The dependence of the decay of correlations on $\sigma^2$ in a
closed rectangular billiard. In each case the dotted lines show the
same quantities as the solid lines, but computed using half number
of discretisation points for both $n$ and $N$.}
\label{EscRate}\end{figure}

Figure \ref{EscRate} (a)  shows a plot of the asymptotic escape rate
$\gamma$ against $\sigma^2$, where the escape rate is given by minus
the logarithm of the spectral radius of the (numerical approximation
to the) operator $\mathcal{L}_{\sigma}$. In each case the opening is
on edge 2, and the two plots shown are for openings of size $0.05$
(from $y=0.1$ to $0.15$) and $0.1$ (from $y=0.15$ to $0.25$). For
large $\sigma$ values we see $\gamma$ settling down to a constant,
the size of which is approximately proportional to the opening size.
This would be expected, since for chaotic maps the asymptotic escape
rate due to a small opening is an exponential decay which to leading
order is proportional to the hole size (see for example \cite{OG12},
\cite{EA13}). For small sigma values we see that the escape rate
decreases towards zero. Again, this reflects the supporting theory
since as the map approaches a deterministic billiard map in a
rectangle, the integrable dynamics and ``sticky" trajectories (small
perturbations of the bouncing ball modes) slow the decay to an
algebraic rate \cite{GZ02}. Such a decay would be reflected by
$\mathcal{L}_{\sigma}$ having a spectral radius of $1$, and hence
$\gamma\rightarrow0$.

Figure \ref{EscRate} (b) shows a plot of the correlation decay rate
$\nu$ against $\sigma^2$, which may also be estimated from the
spectrum of the operator $\mathcal{L}_{\sigma}$. In this case we
look at the size of the second largest eigenvalue $\lambda^*$ of the
closed billiard (the largest eigenvalue is always one for a closed
system). The plot shows
$\nu=-\log(|\lambda^*|)$ increasing with $\sigma^2$. For very small
$\sigma^2$ the plot shows an almost zero decay rate as would be
expected for a system with deterministic and regular dynamics. For
large $\sigma^2$ we see convergence to a value of just over $0.5$,
which clearly indicates the stochastic behaviour introduced from the
noise in the billiard flow. In fact, the dependence of the decay
rate on $\sigma$ appears to follow two distinct behaviours. For
$\sigma^2<0.1$ one sees a rapid increase of $\nu$ with $\sigma$, and
for $\sigma^2>0.1$ the rate of increase is far slower. This can
perhaps be attributed to the PDF governing the noise in the billiard
flow. For $\sigma^2<0.1$, the noise added to the flow is closer to a
non-correlated Gaussian distribution and for $\sigma^2>0.1$, the
scaling and shifting become increasingly significant and the model
approaches a uniform distribution.

Considering Figures \ref{EscRate} (a) and (b) together, a change of
behaviour in the escape rate is also evident close to
$\sigma^2=0.1$. Here the escape rate begins to increase more quickly
before peaking just below $\sigma=1$, and then decreasing to a
constant rate for $\sigma^2>10$. The behaviour for $0.1<\sigma^2<10$
indicates a transition region where the trajectory flow is not yet effectively
random (uniformly distributed), but is also not behaving as a flow
with uncorrelated Gaussian noise. The dotted lines in each case show
a lower precision computation with $n$ and $N$ both halved. The
similarities between the plots suggest a good level of convergence in
the computations. This serves to highlight a further advantage of working
with $\mathcal{L}_{\boldsymbol{\sigma}}$ rather than the FP operator,
where such computations typically show little evidence of convergence \cite{CT11}.

\section{Discussion and conclusions}

A new boundary integral model to propagate ray densities via an
uncertain trajectory flow has been presented. The resulting
phase-space boundary integral representation reduces the
dimensionality of the model, and was shown to directly interpolate
between a deterministic and a random trajectory flow. The model was
implemented numerically via a simple discretisation approach using
piecewise constant collocation in space and a Nystr\"{o}m method in
the momentum variable. Discrete flow mapping type methods were
applied to give a highly efficient computational procedure. An
application to uncertain billiard dynamics in an integrable
rectangular domain was presented; the numerical results demonstrated
the transition between a deterministic and a random flow. Using the
rate of escape and the decay of correlations to further diagnose the
behaviour of the model gave parameter ranges where the model was
effectively behaving as a deterministic trajectory flow with a small
amount of uncorrelated Gaussian noise, a random (uniformly
distributed) flow and a transition phase in between.

In the future, the framework will be extended to three dimensional billiards
by introducing the analog of the PDF (\ref{ScaledPDF}) on
the boundary surface and its corresponding hemispherical momentum
space (see \cite{CTG12}). Practically one would have to also define
an efficient discretisation scheme, but in principle similar methods
to those here can be employed provided the closed boundary surface
consists of (or can be well approximated by) a union of flat
surfaces joined together at their edges. Such an extension would be
important for applications in room acoustics.

A further natural extension arises since one could allow the
parameters $\boldsymbol{\sigma}$ to depend on the phase space
coordinate. In fact, since the PDF (\ref{ScaledPDF}) already depends
on the phase space point indirectly through dependence on $X^{\pm}$,
this extension could be implemented directly in the model here
without extra modification. On a practical note, the dependence of
$\sigma_1$ on the spatial coordinate should to be assumed to be
piecewise constant to match the collocation scheme and maintain the
tractability of the integrals appearing in
(\ref{StochasticTOdisc2}). This extension would be important for
applications in computer graphics, where reflections may take place
from surfaces with different properties. A further consideration
here is that the methods also extend directly to built-up
multi-component structures in the same way as DEA \cite{CGT11}. This
opens up the formulation to applications to built-up vibro-acoustic
structures and complex electromagnetic environments.

\section*{Acknowledgement}
Support from the EU (FP7 IAPP grant MHiVec) is gratefully
acknowledged. We also wish to thank Dr Alex Bespalov and Dr Gabriele Gradoni for
stimulating discussions.


\begin{thebibliography}{99}

\bibitem{CCMM04} A. Celani, M. Cencini, A. Mazzino and M. Vergassola, ``Active and passive fields face
to face," New Journal of Physics \textbf{6} 72 (2004).

\bibitem{SR10} M. Sommer and S. Reich, ``Phase-space volume conservation under space and time discretization
schemes for the shallow-water equations," Monthly Weather Review,
\textbf{138}, 4229-4236 (2010).

\bibitem{Noe09} F. No\'e, C. Sch\"utte, E. Vanden-Eijnden, L. Reich and T.R. Weikl, ``Constructing
the equilibrium ensemble of folding pathways from short
off-equilibrium simulations," Proceedings of the National Academy of
Sciences of the USA \textbf{106}, 19011-19016 (2009).

\bibitem{JK86} J.T. Kayija, ``The rendering
equation," in Proc. SIGGRAPH 1986: 143, DOI:10.1145/15922.15902
(1986).

\bibitem{Vor89} M. Vorl\"ander,
``Simulation of the transient and steady-state sound propagation in
rooms using a new combined ray-tracing/image-source algorithm," J.
Acoust Soc. Am. \textbf{86}, 172-178 (1989).

\bibitem{Cer01} V. \v{C}erven\'y,
\emph{Seismic ray theory} (Cambridge University Press, Cambridge,
UK, 2001).

\bibitem{LeV92} R.J. LeVeque
\emph{Numerical Methods for Conservation Laws}, Lectures in
Mathematics: ETH Z\"urich (Birkh\"auser, Basel, Swizerland, 1992).

\bibitem{Cvi12} P. Cvitanovi\'c, R. Artuso, R. Mainieri, G. Tanner and G. Vattay
{\em Chaos: Classical and Quantum}, {\tt ChaosBook.org} (Niels Bohr
Institute, Copenhagen, Denmark, 2012).

\bibitem{JD96} J. Ding and A. Zhou, ``Finite approximations of Frobenius-Perron operators: A solution of
Ulam's conjecture to multi-dimensional transformations," Physica D
\textbf{92}, 61-68 (1996).

\bibitem{JK09} O. Junge and P. Koltai, ``Discretization of the Frobenius-Perron operator using a sparse Haar
tensor basis - the Sparse Ulam method," SIAM J. Num. Anal.
\textbf{47}, 3464-3485 (2009).

\bibitem{FJK11} G. Froyland, O.
Junge and P. Koltai, ``Estimating long term behavior of flows
without trajectory integration: the infinitesimal generator
approach," SIAM J. Num. Anal. \textbf{51}(1), 223-247, (2013).

\bibitem{BMM12} M. Budisic\', R. Mohrand I. Mezic\',
``Applied Koopmanism," Chaos \textbf{22}, 047510 (2012).

\bibitem{PC10} D. Lippolis and
P. Cvitanovi\'c, ``How well can one resolve the state space of a
chaotic map?" Phys. Rev. Lett., \textbf{104}, 014101 (2010).

\bibitem{BM01} C.J. Bose and R. Murray, ``The exact rate of
approximation in Ulam's method," Discrete and Continuous Dynamical
Systems \textbf{7} 219-235 (2001).

\bibitem{BKL02} M. Blank, G. Keller
and C. Liverani, ``Ruelle-Perron-Frobenius spectrum for Anosov
maps," Nonlinearity \textbf{15} 1905-1973 (2002).

\bibitem{AL06} A. Le Bot, ``Energy exchange
in uncorrelated ray fields of vibroacoustics," J. Acoust. Soc. Am.
\textbf{120}(3), 1194-1208 (2006).

\bibitem{GT09} G. Tanner, ``Dynamical energy
analysis - Determining wave energy distributions in vibro-acoustical
structures in the high-frequency regime," J. Sound. Vib.
\textbf{320}, 1023-1038 (2009).

\bibitem{SAE14} G. Tanner,
D.J. Chappell, D. L\"ochel and N. S\o ndergaard, ``Discrete Flow
Mapping: a mesh based simulation tool for mid-to-high frequency
vibro-acoustic excitation of complex automotive structures," SAE
Int. J. Passeng. Cars - Mech. Syst. \textbf{7}(3) 2014-01-2079
(2014).

\bibitem{CGT11} D.J. Chappell,
S. Giani and G. Tanner, ``Dynamical energy analysis for built-up
acoustic systems at high frequencies," J. Acoust. Soc. Am.
\textbf{130}(3), 1420-1429 (2011).

\bibitem{CTG12} D.J. Chappell,
G. Tanner and S. Giani, ``Boundary element dynamical energy
analysis: a versatile high-frequency method suitable for two or
three dimensional problems," J. Comp. Phys. \textbf{231}, 6181-6191
(2012).

\bibitem{CT13} D.J. Chappell and
G. Tanner, ``Solving the Liouville Equation via a boundary element
method," J. Comp. Phys. \textbf{234}, 487-498 (2013).

\bibitem{PRSL13} D.J. Chappell,
G. Tanner, D. L\"ochel and N. S\o ndergaard, ``Discrete flow
mapping: Transport of ray densities on triangulated surfaces,"
Proc.\ R.\ Soc.\ A, \textbf{469}, 20130153 (2013).

\bibitem{Wamot14} D.J. Chappell,
D. L\"ochel, N. S\o ndergaard and G. Tanner, ``Dynamical energy
analysis on mesh grids: a new tool for describing the vibro-acoustic
response of engineering structures," Wave Motion \textbf{51}(4),
589-597 (2014).

\bibitem{PC98} P. Cvitanovi\'c, C.P. Dettmann, R. Mainieri, and G. Vattay, ``Trace formulas
for stochastic evolution operators: weak noise perturbation theory,"
J. Stat. Phys., \textbf{93}, 981-999 (1998).

\bibitem{PC99a} P. Cvitanovi\'c, C.P. Dettmann, R. Mainieri and G. Vattay,
``Trace formulas for stochastic evolution operators: Smooth
conjugation method," Nonlinearity \textbf{12}, 939-953 (1999).

\bibitem{PC99b} P. Cvitanovi\'c, N. S\o ndergaard, G. Palla,
G. Vattay and C.P. Dettmann,  ``Spectrum of stochastic evolution
operators: Local matrix representation approach,"  Phys. Rev. E
\textbf{60}, 3936-3941 (1999).

\bibitem{GP01} G. Palla, G. Vattay,  A. Voros,  N. S\o ndergaard and C.P. Dettmann, ``Noise corrections to
stochastic trace formulas," Found. Phys. \textbf{31}, 641-657
(2001).

\bibitem{AL98} A. Le Bot, ``A vibroacoustic
model for high frequency analysis," J. Sound. Vib. \textbf{211},
537-554 (1998).

\bibitem{YS04} Ya. G. Sinai, ``What is a billiard?"
Not. Am. Math. Soc., \textbf{51}, 412-413 (2004).

\bibitem{RL69} R.H. Lyon, ``Statistical
analysis of power injection and response in structures and rooms,"
J. Acoust. Soc. Am. \textbf{45}, 545-565 (1969).

\bibitem{RL95} R.H. Lyon and R.G. DeJong, \emph{Theory and Application of Statistical Energy Analysis}, 2nd edn.,
(Butterworth-Heinemann, Boston, USA, 1995).

\bibitem{RCM} S. Hemmady,
T. M. Antonsen Jr., E. Ott and S. M. Anlage, ``Statistical
Prediction and Measurement of Induced Voltages on Components within
Complicated Enclosures: A Wave-Chaotic Approach," IEEE Trans.
Electromagnetic Compatibility, \textbf{54}(4), 758 - 771 (2012).

\bibitem{Gradoni14} G. Gradoni, J-H. Yeh, B. Xiao, T. M. Antonsen, S. M. Anlage and E.
Ott, ``Predicting the statistics of wave transport through chaotic
cavities by the random coupling model: A review and recent
progress," Wave Motion, \textbf{51}(4), 606 - 621 (2014).

\bibitem{RL92} R.S. Langley, ``A wave
intensity technique for the analysis of high frequency vibrations,"
J. Sound. Vib. \textbf{159}, 483-502 (1992).

\bibitem{RL94} R.S. Langley and A.N. Bercin, ``Wave intensity analysis for high frequency
vibrations," Phil. Trans. Roy. Soc. Lond. A \textbf{346}, 489-499
(1994).

\bibitem{AL02} A. Le Bot, ``Energy transfer
for high frequencies in built-up structures," J. Sound. Vib.
\textbf{250}, 247-275 (2002).

\bibitem{CT11} D.J. Chappell and G. Tanner, ``Estimating the validity of statistical energy
analysis using dynamical energy analysis: a preliminary study," in
\emph{Integral Methods in Science and Engineering}, Editors:
Constanda, C. \& Harris, P.J., Birkh\"{a}user, Boston 69-78 (2011).

\bibitem{OG12} O. Georgiou, C.P Dettmann and E.G. Altmann, ``Faster than expected escape
for a class of fully chaotic maps," Chaos, \textbf{22}, 043115
(2012).

\bibitem{EA13} E.G. Altmann, J.S.E. Portela and T. T\'{e}l, ``Leaking Chaotic
Systems," Reviews of Modern Physics \textbf{85} 869-918 (2013).

\bibitem{GZ02} G.M. Zaslavsky,  ``Chaos,
fractional kinetics, and anomalous transport," Phys. Rep.
\textbf{371}(6), 461-580 (2002).
\end{thebibliography}
\end{document}